\documentclass{aa}  

\usepackage{graphicx}
\usepackage{subfigure}
\usepackage{subfloat}
\usepackage{float}
\usepackage[version=3]{mhchem}
\usepackage{booktabs}
\usepackage{fixmath}
\usepackage{pdflscape}
\usepackage{supertabular}
\usepackage{longtable,lscape}
\usepackage[flushleft]{threeparttable}
\usepackage{txfonts}
\begin{document} 

\title{Bright C$_2$H emission in protoplanetary disks in Lupus: high volatile C/O>1 ratios}


\author{A. Miotello\inst{\ref{inst:ESO}}  \and
	    S. Facchini\inst{\ref{inst:ESO}} \and	    
            E. F. van Dishoeck\inst{\ref{inst:leiden},\ref{inst:mpe}}\and 
	    P. Cazzoletti\inst{\ref{inst:mpe}} \and         
	    L. Testi\inst{\ref{inst:ESO}}  \and
            J. P. Williams\inst{\ref{inst:IfA}}\and        
            M. Ansdell\inst{\ref{inst:UCB},\ref{inst:NY}}\\and
            S. van Terwisga\inst{\ref{inst:leiden}}\and    
            N. van der Marel\inst{\ref{inst:canada}}}                

\institute{
European Southern Observatory, Karl-Schwarzschild-Str 2, D-85748 Garching, Germany\label{inst:ESO}\and
Leiden Observatory, Leiden University, Niels Bohrweg 2, NL-2333 CA Leiden, The Netherlands\label{inst:leiden}\and
Max-Planck-institute f{\"u}r extraterrestrische Physik, Giessenbachstra{\ss}e, D-85748 Garching, Germany\label{inst:mpe}\and
Institute for Astronomy, University of Hawaii, 2680 Woodlawn dr., 96822 Honolulu HI, USA\label{inst:IfA}\and
Center for Integrative Planetary Science, University of California at Berkeley, Berkeley, CA 94720, USA\label{inst:UCB}\and
Center for Computational Mathematics \& Center for Computational Astrophysics, Flatiron Institute,162 Fifth Ave New York, NY 10010, USA\label{inst:NY}\and
Herzberg Astronomy \& Astrophysics Programs, National Research Council of Canada, 5071 West Saanich Road, Victoria BC V9E 2E7, Canada\label{inst:canada}}
                       

\abstract{Recent ALMA surveys in different star-forming regions have shown that CO emission in protoplanetary disks is much fainter than expected. Accordingly, CO-based gas masses and gas/dust ratios are orders of magnitude lower than previously thought. This may be explained either as fast gas dispersal, or as chemical evolution and locking up of volatiles in larger bodies leading to the low observed CO fluxes. The latter processes lead to enhanced C/O ratios in the gas, which may be reflected in enhanced abundances of carbon-bearing molecules like C$_2$H. } 
{The goal of this work is to employ C$_2$H observations to understand whether low CO fluxes are caused by volatile depletion, or by fast gas dissipation. } 
{We present ALMA Cycle 4 C$_2$H ($N=3-2$, $J=7/2-5/2$, $F=4-3$ and $F=3-2$) observations of a subsample of nine sources in the Lupus star-forming region. The integrated C$_2$H emission is determined and compared to previous CO isotopologue observations and physical-chemical model predictions.} 
{Seven out of nine disks are detected in C$_2$H, whose line emission is almost as bright as $^{13}$CO. All detections are significantly brighter than the typical sensitivity of the observations, hinting at a bimodal distribution of the C$_2$H line intensities. This conclusion is strengthened when our observations are compared with additional C$_2$H observations of other disks. When compared with physical-chemical models run with DALI, the observed C$_2$H fluxes can be reproduced only if some level of volatile carbon and oxygen depletion is allowed and [C]/[O]>1 in the gas. Models with reduced gas/dust ratios near unity fail instead to reproduce the observed C$_2$H line luminosity. A steeper than linear correlation between C$_2$H  and CN emission line is found for the Lupus disks. This is linked to the fact that C$_2$H emission lines are affected more strongly by [C]/[O] variations than CN lines. Ring-like structures are detected both in C$_2$H and in continuum emission but, as for CN, they do not seem to be connected. Sz 71 shows ring shaped emission in both C$_2$H and CN with the location of the peak intensity coinciding, within our 30 au resolution.} 
{Our new ALMA C$_2$H observations favour volatile carbon and oxygen depletion rather than fast gas dispersal to explain the faint CO observations for most of the disks. This result has implications for disk-evolution and planet-formation theories, as disk gas masses may be larger than expected if CO is considered to be the main carbon carrier in the gas phase.} 


\maketitle
%
\section{Introduction}
\begin{figure*}[h]
   \resizebox{\hsize}{!}
             {\includegraphics[width=\textwidth]{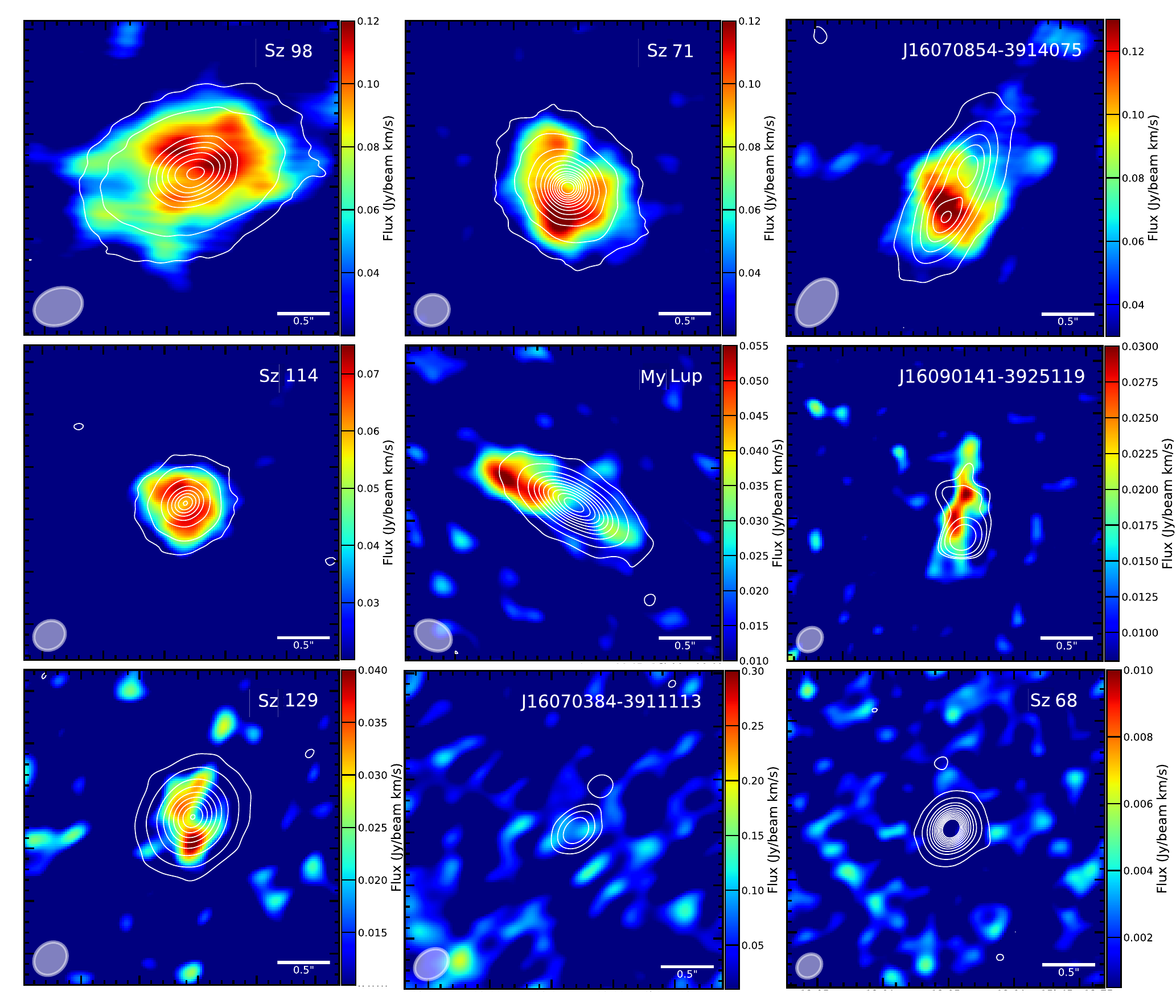}}
      \caption{C$_2$H ($N=3-2, J=7/2-5/2, F=4-3$) moment 0 maps of the nine Lupus sources obtained integrating the emission over the channels where signal was detected. No Keplerian masking was used to produce such maps. White contours show the continuum emission at 3$\sigma$, 20$\sigma$ and at 20$\sigma$ intervals.}
       \label{mom_0}
\end{figure*}
Protoplanetary disks are studied since decades and thousands of exoplanets have been discovered and characterised in the last 20 years. Yet, how disks evolve and lead to the formation of such planets is unclear. In particular, disk parameters, such as their surface density distribution and their vertical structure, and even more fundamental properties such as their total mass and the gas-to-dust mass ratio, are poorly constrained by observations. 

Large mm-sized grains located in the disk midplane, which account for the bulk of the dust mass, are usually traced by their thermal continuum emission at (sub-)mm wavelengths. However, significant assumptions on the dust opacity, on its temperature, and on the optical depth of the continuum emission have to be made, in order to obtain the total dust mass \citep[see e.g,][]{Andrews13,Testi14,Pascucci16,Tripathi17,Ansdell18}. 

The bulk of the gas is typically traced by carbon monoxide's (CO) less abundant isotopologue emission through pure rotational lines, which originate in the warm molecular layer of the disk. Being the second most abundant molecule after molecular hydrogen (H$_2$), and having a well known and relatively simple chemistry, CO is typically used to trace the bulk of the gas in disks. More precisely, $^{13}$CO and C$^{18}$O emission is less optically thick and best suited to get information about the column densities and to derive the disk total mass \citep{vanZadelhoff01,Williams14}. 
Freeze-out and isotope-selective photodissociation, the main processes governing the abundance of CO in disks, are implemented  with various levels of detail in different physical-chemical disk models \citep[e.g.,][]{Visser09,Williams14,Miotello14,Woitke16} and properly considered in the data analysis. Additionally, a big assumption regarding the disk volatile [C]/[H] ratio is needed to convert CO mass to H$_2$ mass; typically the [C]/[H] =$1.35\times10^{-4}$ value of the diffuse ISM is used and presumed constant throughout the disk, which results in CO locking up the bulk of the volatile elemental carbon with respect to hydrogen.

Thanks to the advent of the Atacama Large Millimeter/submillimeter Array (ALMA), large surveys of protoplanetary disks in different star forming regions have been carried out to study the gas and dust components simultaneously \citep[][]{Ansdell16,Ansdell17,Barenfeld16, Eisner16, Pascucci16}. A result that is common to these surveys is that CO emission from disks is fainter than expected \citep{Miotello17,Long17}. This may be interpreted as lack of gas due to fast disk dispersal, or as lack of volatile carbon that leads to faint CO lines \citep{Favre13,Ansdell16,Schwarz16,Miotello17}. The second interpretation is in line with previous studies where HD-based mass measurements of a few bright disks (TW Hya, DM Tau and GM Aur) were found to be incompatible with much smaller CO-based disk masses, that assume a normal ISM volatile carbon abundance \citep{Bergin13,Favre13,McClure16}. In particular, it was shown that the TW Hya disk needs to be depleted in volatile carbon by one to two orders of magnitude in order to reproduce the HD and CO isotopologue line emission \citep[][]{Favre13,Kama16,Trapman17}. Furthermore, in order to reproduce hydrocarbon observations, more precisely C$_2$H emission, volatile oxygen needs to be depleted too and by a larger amount than carbon, in order to get to carbon-to-oxygen ratios ([C]/[O]) larger than one \citep{Kastner15,Kama16,Bergin16}. Pre-ALMA observations of C$_2$H were carried out by \cite{Dutrey97} for DM Tau and GG Ta, by \cite{Henning10} for MWC 480, LkCa 15 and DM Tau, and by \cite{Kastner15} for TW Hya. ALMA C$_2$H data have been published for TW Hya, DM Tau, IM Lup and V4046 Sgr.\citep{Bergin16,Cleeves18,Kastner18}. Similarly to TW Hya and DM Tau, volatile carbon and oxygen abundances have been constrained in the IM Lup disk based on the detection of C$_2$H and found to be depleted \citep{Cleeves18}. During revision of this manuscript, \cite{Bergner19} have published a survey of C$_2$H observations of 14 disks observed with ALMA spanning a range of ages, stellar luminosities, and stellar masses, which cover some of the sources mentioned above. 

More generally, it is known that volatile oxygen must be depleted thanks to the weak/non-detections of H$_2$O lines observed by the \emph{Herschel Space Observatory} \citep{Hogerheijde11,Du17}. This implies that volatile oxygen is locked up in larger bodies in the midplane that do not circulate to larger disk heights where the water would be photodesorbed. Carbon depletion may be caused by two different processes. Similarly to oxygen, carbon may be locked up in larger bodies and sequestered in the midplane \citep{Krijt18}. Another option is chemical transformation of carbon, most notably into CO$_2$, CH$_3$OH and hydrocarbons \citep{Eistrup16,Eistrup18,Yu17a,Yu17b,Bosman17,Bosman18,Schwarz18}.

If [C]/[O]>1, some carbon is left free from CO and is available to create more complex molecules, such as hydrocarbons. Moreover, if the process of volatile depletion happens by incorporating carbon and oxygen in larger grains which settle to the midplane and drift inwards, the disk regions where this happens would be more exposed to UV radiation. High carbon-to-oxygen ratios combined with strong UV irradiation are the right conditions to boost hydrocarbon emission \citep{Bergin16}. If the faint CO lines found in protoplanetary disks are due to high levels of volatile depletion rather than to low gas-to-dust ratios, this should be reflected in strong hydrocarbons -- for instance C$_2$H -- fluxes. 

To address these questions, new Cycle 4 ALMA C$_2$H line emission data  have been observed in a subsample of disks in the Lupus star-forming region and are presented in Sec. \ref{observations}. The main results, shown in Sec. \ref{results}, are discussed in Sec. \ref{discussion} where C$_2$H line emission is linked to the volatile [C]/[O] ratio in disks.
\begin{table*}[h]
\footnotesize
\caption{Target list with integrated fluxes and distances.}
\label{Tab1}
\begin{center}
\begin{tabular}{lccccccccc}
\toprule
\toprule
Target&RA&Dec&$M_{\star}$\tnote{1}$^{\rm (a)}$&$L_{\star}$\tnote{2}$^{\rm (a)}$&log $L_{\rm acc}$\tnote{2}$^{\rm (a)}$&$F_{\rm C_2H}$&$F_{890\mu \rm m}$\tnote{3}$^{\rm (b)}$&$F_{\rm^{13}CO}$\tnote{3}$^{\rm (b)}$&$d$\tnote{4}$^{\rm (c)}$\\
&[J2000]&[J2000]&$[M_{\odot}]$&$[L_{\odot}]$&$[L_{\odot}]$&[mJy km s$^{-1}$]&[mJy]&[mJy km s$^{-1}$]&[pc]\\
\cmidrule(lr){1-10}
Sz 98&16:08:22.481&-39:04:46.812&0.67&1.53&-0.71&1101$\pm$57&237.29$\pm$0.51&506$\pm$120&156.2\\	
Sz 71&15:46:44.716 &-34:30:36.057&0.41&0.33&-2.17&1101$\pm$46&166.04$\pm$0.37&1298$\pm$107&155.9\\	
J16070854-3914075	&16:07:08.539&-39:14:07.885&-&-&-&647$\pm$55&92.07$\pm$0.49&1232$\pm$173&175.8\\
Sz 114&16:09:01.836 &-39:05:12.790&0.19&0.21&-2.68&383$\pm$32& 96.41$\pm$0.50&716$\pm$50&162.2\\
MY Lup&16:00:44.503 &-41:55:31.271&1.09&0.85&-1.00&284$\pm$21&176.81$\pm$0.56&874$\pm$54&156.6\\	
J16090141-3925119&16:09:01.405 &-39:25:12.345&0.2&0.07&-1.5&200$\pm$46&17.50$\pm$0.68&1637$\pm$89&164.3\\
Sz 129&15:59:16.458 &-41:57:10.662&0.78&0.43&-1.13&107$\pm$16&181.12$\pm$0.52&636$\pm$80&161.7\\
J16070384-3911113&16:07:03.825 &-39:11:11.756&-&0.003&-5.40&<16.1$^{\rm (d)}$&4.52$\pm$0.55&1242$\pm$98&150.0\\
Sz 68&15:45:12.848&-34:17:30.986&-&5.42&1.18&<5.3$^{\rm (d)}$&150.37$\pm$0.46&915$\pm$131&154.2\\
\hline
\end{tabular}
\end{center}
 \begin{tablenotes}
    \item[1] (a) Stellar properties corrected after Gaia DR2 (Alcala et al., under revision). (b): \cite{Ansdell16}. (c): Distance measured by GAIA \citep{GAIA}. Typical errors on the distance are 1-2\% of the measured distance for these sources. (d): 3$\sigma$ upper limits.
  \end{tablenotes}
\end{table*}
\section{Observations}
\label{observations}

Observations of ten protoplanetary disks in the Lupus star-forming region were acquired with ALMA in Cycle 4, under the project code 2017.1.00495.S (PI: Miotello A.). The targets were selected among the brightest Lupus disks in continuum emission as those detected in CO isotopologues and in CN \citep{Ansdell18,vanTerwisga18} and with CO-based gas-to-dust ratios ranging from 100 to 0.1 (see Table \ref{Tab1}). Six sources, J16011549-4152351, J16070854-3914075, MY Lup, Sz 114, Sz 71 and Sz 98, were observed on 2017 July 12 with 49 antennas  (16.7-2700 m baselines) using J1517-2422 and J1610-3958 as calibrators. Four targets, J16070384-3911113, J16090141-3925119, Sz 129 andSz 68, were observed on 2017 July 15 with 42 antennas (18.6-1500 m baselines) using J1427-4206 and J1610-3958 as calibrators. The typical integration time was approximatively 20 minutes on source. We estimate a flux calibration error of 10\% based on variations in the flux calibrators.

The spectral setup included five windows covering C$_2$H, c-C$_3$H$_2$, SO, H$^{13}$CO$^+$, and H$_2$CO lines, and one spectral window devoted to the continuum emission centred on 246 GHz. Here, together with the continuum data, we present only the C$_2$H ($N=3-2, J=7/2-5/2, F=4-3$ and $F=3-2$) line observations, whose spectral window was centred on $262.006$ GHz, with bandwidth of 117 MHz, channel width of 0.14 MHz, and velocity resolution of 0.16 km s$^{-1}$. Only two hyperfine components ($F=4-3$ and $F=3-2$) of the C$_2$H ($N=3-2, J=7/2-5/2$) line are covered. No clear detection of the other molecules is found in any of the targeted sources.

From the pipeline calibrated data we performed (only phase and not amplitude) self-calibration using each source's continuum emission. The data were imaged using the \texttt{clean} task in CASA version 4.7.2 using natural weighting and the achieved beam size is $\sim$ 0.2 $\times$ 0.3 arcseconds. Images are presented in Fig. \ref{mom_0}. The typical achieved rms is between 5 and 10 mJy km s$^{-1}$. Integrated fluxes, with the achieved rms are reported for each source in Table  \ref{Tab1}, together with $^{13}$CO integrated fluxes and continuum flux densities, taken from \cite{Ansdell16}.

Similarly to what was found in the $^{12}$CO data, the C$_2$H emission in J16011549-4152351 is completely contaminated by the foreground cloud. Therefore, it is impossible to extract any meaningful information on the disk's hydrocarbon emission. For this reason J16011549-4152351 is excluded from the analysis presented in this paper.

\begin{figure}[h]
   \resizebox{\hsize}{!}
             {\includegraphics[width=2\textwidth]{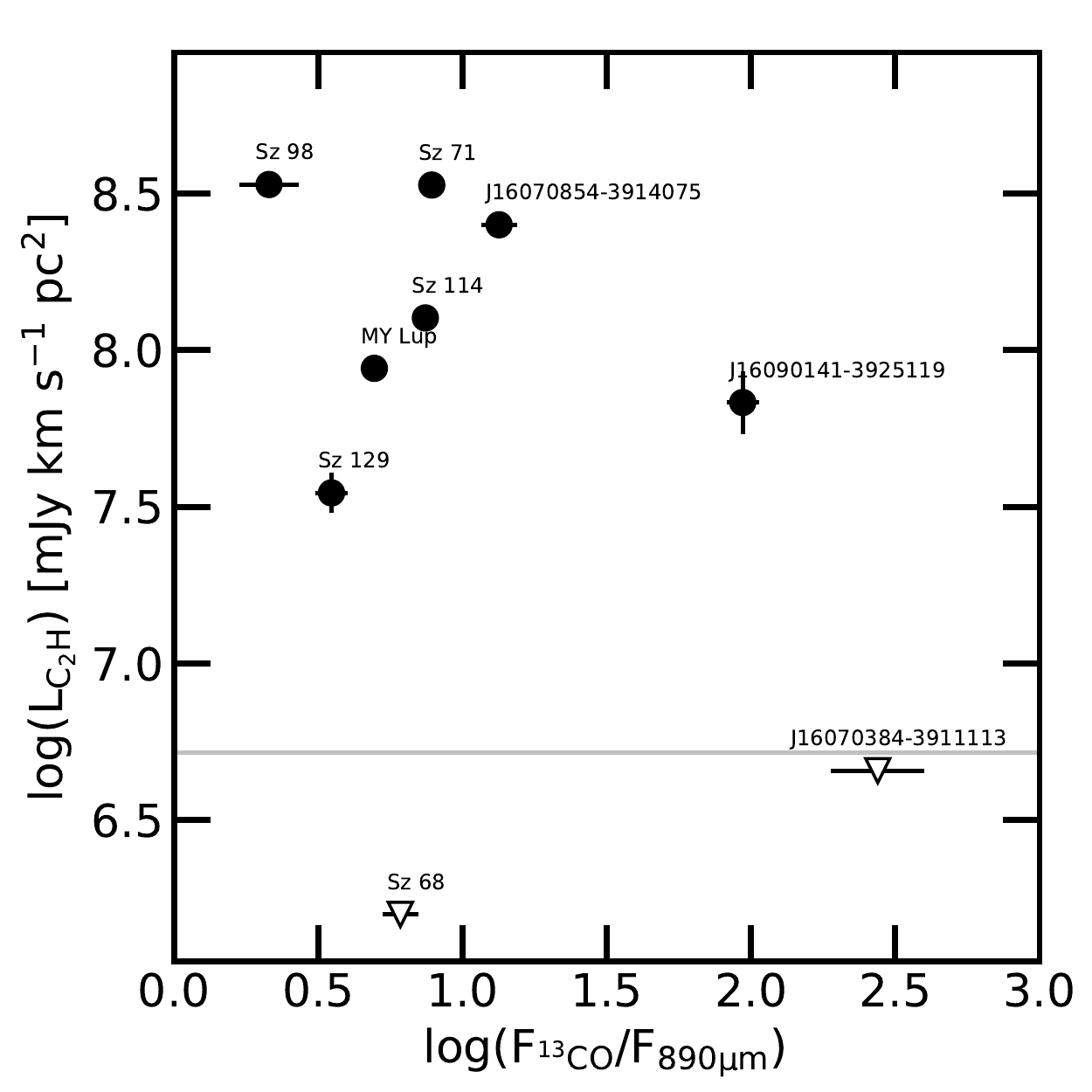}}
      \caption{Logarithm of C$_2$H ($N=3-2, J=7/2-5/2, F=4-3$ and $F=3-2$) integrated line luminosity of nine Lupus protoplanetary disks versus the logarithm of $^{13}$CO (\emph{J}=3-2)integrated line flux over the continuum total flux density at 890 $\mu$m (integrated over a band width of 1 km s$^{-1}$ for convenience). The solid grey line shows the typical sensitivity obtained at the distance of our lowest detection (Sz129, at 162 pc). }
       \label{obs}
\end{figure}
\section{Results}
\label{results}

\subsection{Integrated fluxes}

Seven out of nine disks in our sample are detected in C$_2$H emission, which is generally bright, of the same order of $^{13}$CO emission (see Table \ref{Tab1}). The moment 0 maps of the  C$_2$H line emission in our targets are presented in Figure \ref{mom_0}. These maps are obtained by integrating the emission from all channels where signal was detected. The disk integrated C$_2$H fluxes (reported in Table \ref{Tab1}) are measured on the image plane, using an aperture synthesis method on the moment 0 maps, similar to that used by \cite{Ansdell16}. The line luminosity is obtained by multiplying the integrated flux by $4 \pi d^2$, where $d$ is the distance of the target in parsecs as measured by GAIA \citep[see Table \ref{Tab1},][]{GAIA}. 
The C$_2$H integrated line luminosity is plotted in Fig. \ref{obs} against $^{13}$CO over continuum at 890 $\mu$m fluxes, where the continuum total flux density is integrated over a band width of 1 km s$^{-1}$ for convenience \citep[Table \ref{Tab1},][]{Ansdell16}. The  $^{13}$CO-to-continuum flux ratio can be used as an observational proxy of the gas/dust mass ratio, under the assumption that CO is the main carbon carrier in the disk warm layer. The typical sensitivity calculated at 162 pc, the distance of Sz129 which is our lowest detection, is shown by the grey solid line. A bimodal distribution is observed: either C$_2$H emission is strongly detected at more than a 5$\sigma$ level, or such molecule is not detected. A possible interpretation of this behaviour is discussed in Sec. \ref{discussion}.
C$_2$H emission is likely optically thin for the observed sources in Lupus, as later discussed.

\subsection{Ring-like structures}
ALMA is revolutionising the field of planet-formation by showing that disks often present substructures, almost always in the shape of cavities \citep[e.g.,][]{vanderMarel16}, rings and gaps \citep[see][and references therein]{Long18,Andrews18}. Such structures, seen in continuum emission, are possibly related to perturbations of the disk gas surface density distribution, due to the presence of forming planets, or to different suites of instabilities \citep[e.g.,][]{Johansen09,Bai14,Flock15}. Rings have also been detected in molecular line emission, but are of a different nature \citep[see e.g.,][]{Bergin16,Cazzoletti18,Terwisga18}. The cause of such rings varies for different molecules, but is related to chemistry, rather than a physical perturbation of the disk structure. Rings seen in DCO$^+$ and N$_2$H$^+$ are, for example, possibly linked to the location ice rings, such as that caused by the CO snow-surface \citep[e.g.,][]{Mathews13,Oberg15,vtH17}. In other cases, such as for CN and for C$_2$H, there is no relation with the snow lines \citep[][]{Bergin16,Cazzoletti18,vanTerwisga18}. The link between the structures seen in the continuum and in different gas tracers is being investigated with current and upcoming observations. 

\begin{figure}[h]
   \resizebox{\hsize}{!}
             {\includegraphics[width=2\textwidth]{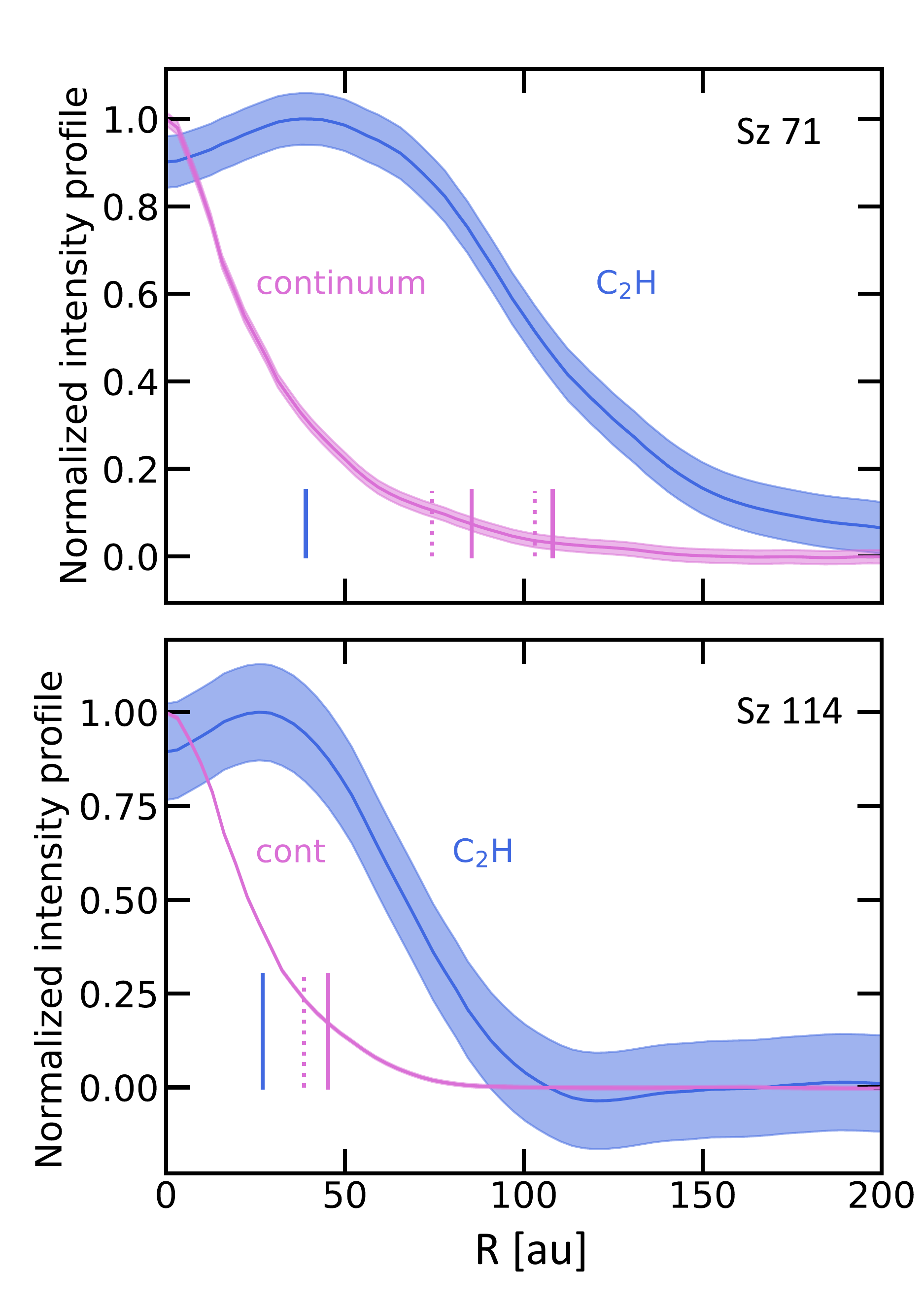}}
      \caption{Azimuthally averaged normalised brightness profiles of the C$_2$H (blue) and 1.2 mm continuum (pink) emission as functions of disk radius for Sz 71 (top panel) and Sz 114 (bottom panel). The blue vertical lines show the peak location of the C$_2$H emission, while the dotted and solid pink lines show respectively the location of the gaps and rings detected in these disks by \cite{Huang18} with higher angular resolution ALMA data \citep[DSHARP large program;][]{Andrews18}.}
       \label{prof}
\end{figure}

Ring-like structures are detected both in continuum emission at 1.2 mm and in C$_2$H line emission also in our sample. Thanks to the much higher sensitivity than the \cite{Ansdell16} data, our ALMA observations allowed us to detect continuum emission at 1.2 mm at high signal-to-noise ratios. The continuum maps of the nine disks presented here are shown by the white contours in Figure \ref{mom_0}. On the image plane, only Sz 98 shows the presence of a ring in continuum emission located at 90 au, as already hinted at by the modelling on the $uv$-plane by \cite{Tazzari17} and by \cite{vanTerwisga18} (see  Fig. \ref{mom_0}). 
Sz 71, Sz 114, and Sz 129 are also included in the ALMA large program "DSHARP" \citep{Andrews18} and, when observed at such high angular resolution, they show ring and gap structures in the outer disks \citep{Huang18}.

As shown in Figure \ref{mom_0}, C$_2$H emission is Sz 114 show clear ring-shaped emission. The C$_2$H azimuthally averaged intensity radial profile is computed after derotating and deprojecting the images following \cite{vanTerwisga18} (see blue line in Fig. \ref{prof}). The azimuthally averaged continuum emission in our observations is shown by the pink line in Fig.  \ref{prof}. The peak of C$_2$H emission is located at 40 au and 27 au respectively in these two sources. This former peak coincides with the CN emission peak found  by \cite{vanTerwisga18} in Sz 71. The location of the gaps and rings detected at very high angular resolution ALMA data ($\sim$30 to 60 mas) by \cite{Huang18} in these two sources is indicated by the dotted and solid pink vertical lines respectively, while the peak of the C$_2$H emission is shown by the blue vertical line. The C$_2$H emission ring location does not seem to be connected with either the gaps or the rings seen in continuum emission in Sz 71. The angular resolution of our C$_2$H observations is 30 au, and this allows us to rule out that the C$_2$H ring location coincides with any of the rings and gaps discovered in the continuum by \cite{Huang18}. This hints to a chemical origin for the hydrocarbon rings, similar to what found for CN at same location \citep[][]{Cazzoletti18,Terwisga18}. For Sz 114 the interpretation is less clear, as the distance between the C$_2$H ring and the continuum substructures is smaller than our angular resolution. Proper chemical modelling to interpret the nature of the observed molecular rings is beyond the scope of this paper.

\section{Discussion}
\label{discussion}

Following \cite{Bergin16} and \cite{Kama16}, if depletion of volatile carbon and oxygen happens in such a way that [C]/[O]>1 in the gas, disks with lower carbon abundance and fainter CO isotopologue lines would show brighter hydrocarbon lines. If [C]/[O]>1, not all volatile carbon is locked in CO but some is available to start the hydrocarbon chemistry. Furthermore, if the process of volatile depletion happens by incorporating carbon and oxygen in larger grains which settle to the midplane and drift inwards, the disk regions where this happens would be more exposed to UV radiation. High carbon-to-oxygen ratios combined with strong UV irradiation are the right conditions to boost hydrocarbon emission \citep{Bergin16}. So, if the faint CO lines found in Lupus disks are due to high levels of volatile depletion rather than to low g/d ratios, this should be reflected in strong hydrocarbon emission. It may be expected that sources with volatile C and O depletion and thus faint CO lines will show bright C$_2$H lines and vice versa.

\subsection{CO-based g/d ratios}

The $^{13}$CO and C$^{18}$O integrated fluxes and continuum flux densities from \cite{Ansdell16} can be used to estimate CO-based gas and dust masses. \cite{Miotello17} have employed the grid of physical-chemical models run with the code DALI \citep{Bruderer12} presented previously in \cite{Miotello16}, which include isotope-selective processes and freeze-out, to estimate CO-based gas masses for the disks in Lupus. Such models cover a realistic range of disk and stellar parameters, sample disk masses from $10^{-5} M_{\odot}$ to  $10^{-1} M_{\odot}$, and produce simulated spatially integrated continuum and line fluxes. For each mass bin, \cite{Miotello17} computed a median integrated $^{13}$CO line luminosity and they compared such values with observations \citep{Ansdell16}, to determine the disk gas mass which is an input parameter in the DALI models. The uncertainty on the mass estimate was given by the dispersion from models with different disk parameters. The overall  CO-based gas-masses are very low, often lower than one $M_{\rm J}$. 
As a consequence, the typical g/d mass ratios found in Lupus are much lower than the factor of 100 measured in the ISM \citep{Miotello17}. Such low CO-based g/d ratio may in fact be an indication of volatile carbon and oxygen depletion, rather than fast gas dispersal in Lupus disks.

\subsection{Dependence of C$_2$H vs CO-based g/d mass ratio.}

It is a priori not clear how tight any anti-correlation between C$_2$H and CO-based g/d ratio should be, and even if this should be linear. This is in fact probably not the case, as there are competing effects at play that may differ from source to source. If carbon is depleted by a factor of 100, one could naively explain why CO emission is found to be a factor of 100 fainter than expected, but not C$_2$H emission, which should be fainter by a factor of 10$^4$. C$_2$H is instead found to be bright in the studied disks and this may be due to volatile [C]/[O]>1 which boosts the C$_2$H emission irrespective of the low C abundance. How much carbon depletion and [C]/[O]>1 balance each other and which one prevails depends on source details and amount of depletion, but a tight relation is not expected. Any non-correlation or anti-correlation goes already in the direction of [C]/[O]>1 hypothesis.

A simple interpretation of the bimodal distribution seen in Fig.  \ref{obs} could be that the volatile [C]/[O] ratio would play as an on/off switch of the C$_2$H emission. C$_2$H lines are bright only if [C]/[O]>1 and the hydrocarbon chemistry path is allowed, otherwise C$_2$H lines would be much fainter, i.e., undetected in our specific case. This would be in line with chemical models, showing that the C$_2$H column densities are boosted as soon as the [C]/[O] becomes larger than unity \citep[see e.g.,][and Sec. \ref{DALI_sec}]{Cleeves18}. Such a scenario would explain the sharp transition between detections and non-detections seen in Fig. \ref{obs}. All detections in our sample are much brighter than the typical sensitivity of our observations, shown by the grey line in Fig. \ref{obs}. Our interpretation is however limited by the small-number statistics: our sample may simply lack intermediate cases. Enlarging the sample of deep C$_2$H observations in Lupus would allow us to better describe the relation between C$_2$H and CO-based g/d ratio and to learn more about the volatile depletion scenario.

\subsection{C$_2$H emission in other sources}

Recently, \cite{Bergner19} have published C$_2$H observations of a set of 14 disks spanning a range of ages, stellar luminosities, and stellar masses, including the previously observed disks by \cite{Dutrey97}, \cite{Henning10}, \cite{Guilloteau16}, \cite{Bergin16}, \cite{Cleeves18}, and \cite{Kastner18}. This sample is complementary to our set of observations, as we target fainter and less radially extended disks, shown to be more representative of the disk bulk population \citep[see e.g.,][]{Ansdell16,Ansdell18,Pascucci16,Long18}, with similar stellar parameters, and age \citep[Lupus is 1-3 Myr old,][]{Comeron08}. 
\begin{figure*}[h]
   \resizebox{\hsize}{!}
             {\includegraphics[width=2\textwidth]{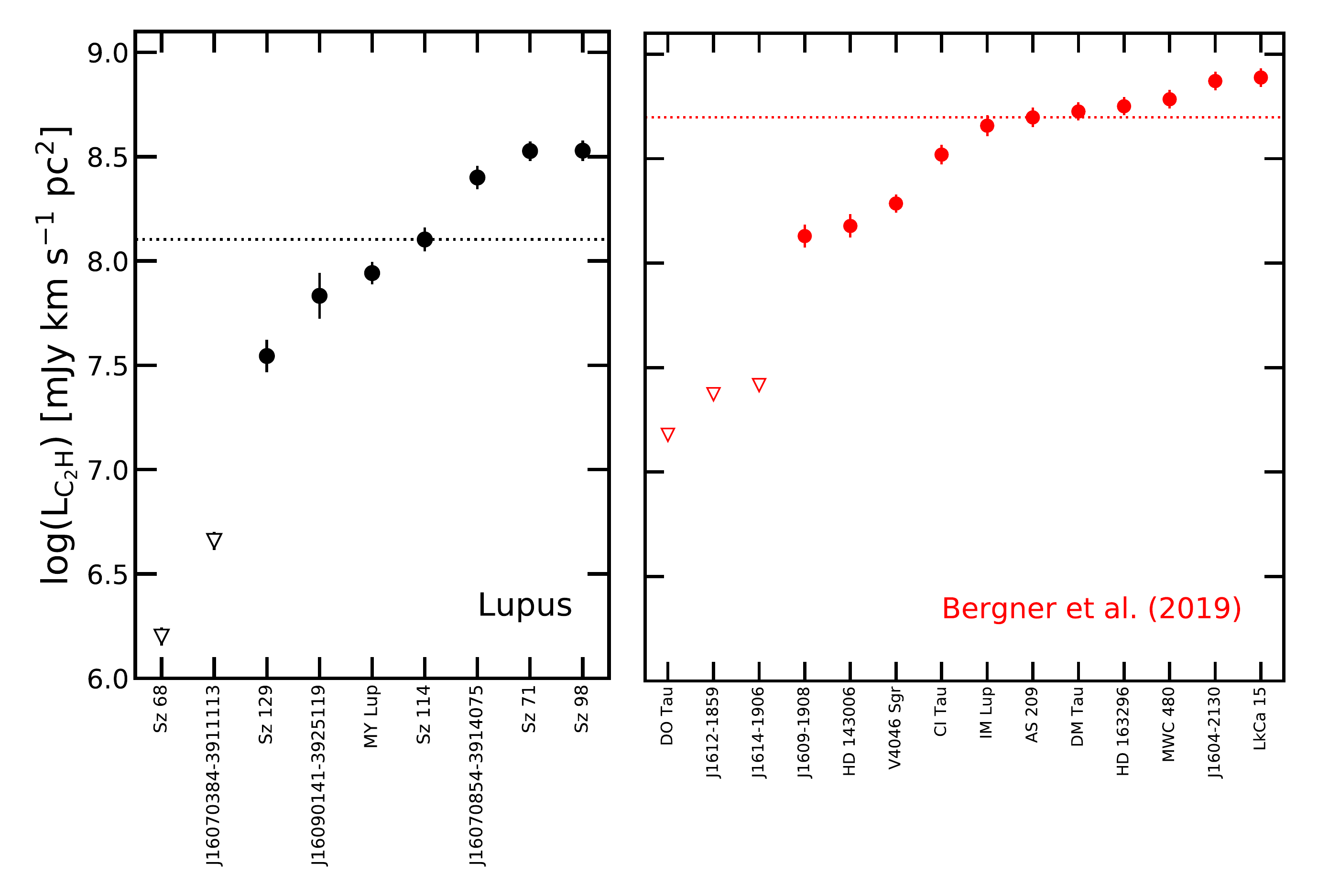}}
      \caption{Logarithm of C$_2$H ($N=3-2, J=7/2-5/2, F=4-3$ and $F=3-2$) integrated line luminosity of the new Lupus observations presented in this paper in black (left panel) and of the disks presented by \cite{Bergner19} in red (right panel). Empty triangles show the C$_2$H non-detections as 3-$\sigma$ upper limits. The dotted lines show the median integrated C$_2$H line luminosity of the detected Lupus disks, in black, and of the detected additional sources, in red. Note the much larger dynamic range of our ALMA data.}
       \label{all}
\end{figure*}
The Lupus disks are shown by the black symbols in Fig. \ref{all}, while the additional sources are reported in red. The C$_2$H non-detections are shown by the empty triangles as 3-$\sigma$ upper limits. 
The observations presented by \cite{Bergner19} have have less sensitivity than ours, with 3$-\sigma$ limits typically a factor 10 higher,  and they do not allow us to rule out the hypothesis of a step-like behaviour of the C$_2$H emission by populating the plot with intermediate cases, i.e., with the logarithm of C$_2$H integrated line luminosities between 6.8 and 7.5. The disks in the \cite{Bergner19} sample that might populate this range of C$_2$H luminosity are all upper limits. On the other hand, our step-like behaviour hypothesis is strengthened by their detections, that are all as bright or brighter than our Lupus observations.

The two samples are however complementary. In Fig. \ref{all} the dotted lines show the median integrated C$_2$H line luminosity of the Lupus disks, in black, and of the additional sources, in red, which is 0.5 dex higher. 
The disks presented by \cite{Bergner19} host in many cases more luminous stars (e.g., HD163296, MWC 480), show brighter millimetre continuum fluxes, and they are also much larger in their radial extent \citep[see Tab. 5, and 6 of ][]{Bergner19}. The UV field impinging onto the disk, which is intrinsically connected with the stellar properties, and the disk outer radius are two parameters that most likely affect the C$_2$H emission, similarly to what found for CN by \citep{Cazzoletti18} and \citep{vanTerwisga18}.

The question we would like to answer with this paper, however, regards the nature of the faint CO isotopologue emission of the Lupus disks \citep{Ansdell16,Miotello17}.  As found also in Chamaeleon, the bulk of the disk population seems to be composed of faint  and compact disks \citep{Pascucci16,Long17}. The open question is whether such disks have low gas masses or low volatile carbon and oxygen abundance, as both scenarios would lead to faint CO fluxes. Disks surrounding brighter stars and with larger radial extent, such as those presented e.g. by \cite{Bergner19}, do not show fainter than expected CO emission \citep[see e.g. HD 163296,][]{Kama16}. A proper description of the C$_2$H emission dependence on higher UV fields and larger disk radii is beyond the scope of this paper, despite it being an interesting and important topic to be explored with models, similarly to what was done for CN by \cite{Cazzoletti18}. For these reasons, we focus here on the Lupus disk observations and we compare them with an explorative grid of models, tapered on such targets (see Sec. \ref{DALI_sec}). 

\subsection{C$_2$H and CN line luminosity in Lupus }

\begin{figure}[h]
   \resizebox{\hsize}{!}
             {\includegraphics[width=2\textwidth]{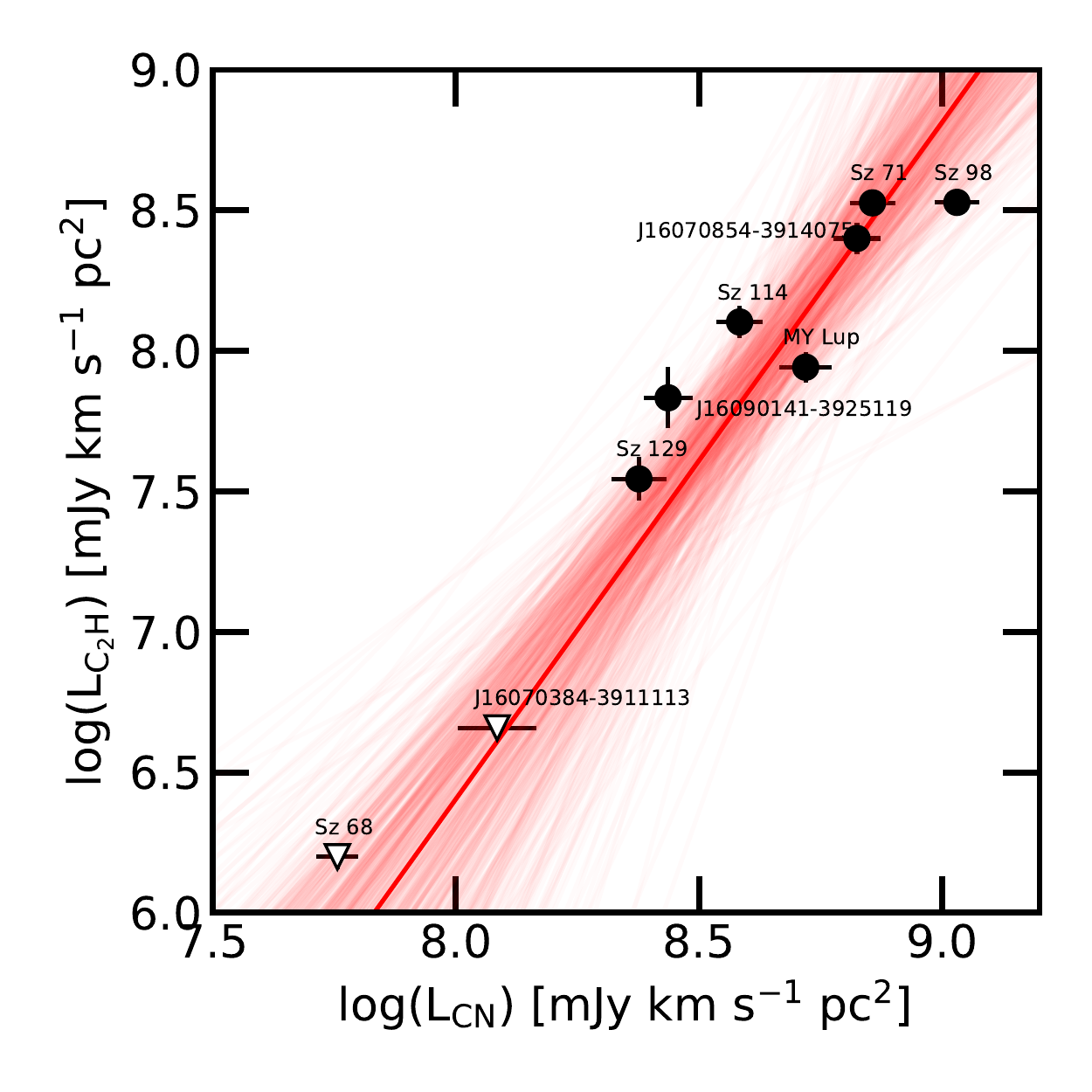}}
      \caption{Logarithm of C$_2$H ($N=3-2, J=7/2-5/2, F=4-3$ and $F=3-2$) versus CN ($N=3-2$) integrated line luminosity \citep{vanTerwisga18} for the Lupus disks. Empty triangles show the C$_2$H non-detections as 3-$\sigma$ upper limits. The solid red line shows the best fit obtained using the Bayesian fitting procedure by \cite{Kelly07}, while the light red lines show a subsample of results from some chains.}
       \label{CN}
\end{figure}

Figure \ref{CN} shows integrated C$_2$H line luminosity for the nine Lupus sources as a function of integrated CN line luminosity. The C$_2$H non detections are also reported and shown as upper limits at the 3-$\sigma$ level by the empty triangles. We compute the linear regression coefficients using the fully Bayesian method by  \cite{Kelly07}, which allows the inclusion of  uncertainties on both axes, upper limits and intrinsic dispersion in the fitting procedure.
    A correlation between the logarithm of C$_2$H and CN line luminosity is found for the Lupus disks. The best representative set of parameters, which we adopt as the median of the results of the chains, is shown in Fig. \ref{CN}  by the red solid line while the light red lines show a subsample of results from some chains. The intercept and the slope of the linear regression are $\alpha= -12.6\pm 4$ and $\beta = 2.4\pm0.4$. The best representative model has a correlation coefficient of $0.96\pm0.04$. As already mentioned, this may be an additional indication that C$_2$H and CN have a similar or common driver in protoplanetary disks, as found by \cite{Bergner19} for C$_2$H and HCN. The linear correlation found on the log-log space is however steeper than two, hinting to the fact that C$_2$H emission must depend stronger than CN emission on some parameter. C$_2$H and CN emission luminosities, however, do not show different trends with stellar or accretion luminosity (see Figures \ref{C2H_Lstar}, \ref{CN_Lstar}, \ref{C2H_Lacc}, \ref{CN_Lacc} in Appendix \ref{appendix}) ruling out a different dependence on the UV field.

\cite{Kastner14} and \cite{Guilloteau16} have shown that C$_2$H and CN abundances may be enhanced in more evolved disks. Following this suggestion, Sz 68 and J16070384-3911113, the two Lupus sources where C$_2$H was not detected, may be the youngest disks in our sample. However, there is no indication from their classification that they are less evolved than the other 7 sources. This may simply be a sign that  Sz 68 and J16070384-3911113 are chemically less evolved. On the other hand, the survey by \cite{Bergner19} was explicitly designed to check for evolutionary trends, as it contains disks with ages from roughly 1 to 13 Myr. The authors find that there is no clear relation between older disks and brighter C$_2$H lines. The caveat to their analysis is that fluxes may not be tracing molecular abundances reliably, due to optical depth effects. This problem should however not apply to our Lupus observations, as the observed fluxes are lower than those found by \cite{Bergner19}.

\subsection{Comparison with models}
\label{DALI_sec}
In order to better quantify the effect of carbon and oxygen depletion on the hydrocarbon line emission, a grid of physical-chemical disk models run with the code DALI \citep[Dust and LInes;][]{Bruderer12} was employed. This is a similar set of models as presented by \cite{Cazzoletti18} for CN and they are designed to cover the typical disk parameter space, in terms of disk mass, vertical structure, and radial extent, with an updated chemistry described by \cite{Visser18}. More specifically, we chose T Tauri like disk models, where the central star emission is described by a black body at $T_{\rm eff}=4000$ K plus an UV excess which mimics a typical mass accretion rate, and $L_{\rm bol}=1 L_{\odot}$. The large-over-small grains mass fraction is $f_{\rm large}=0.99$ and large grains are settled toward the midplane with a settling paramer $\chi=0.2$ (see \cite{Cazzoletti18} for more details). Such models assume typical ISM-like gas-phase elemental abundances, more precisely [C]/[H]=$1\times 10^{-4}$ and [O]/[H]=$3.5\times 10^{-4}$ ([C]/[O]$\sim$0.3).  A detailed source-by-source modelling is beyond the scope of this paper. We employ the aforementioned grid of models to obtain physically and chemically motivated predictions that we can compare with the observed C$_2$H integrated line luminosities.
\begin{figure*}[h]
   \resizebox{\hsize}{!}
             {\includegraphics[width=2\textwidth]{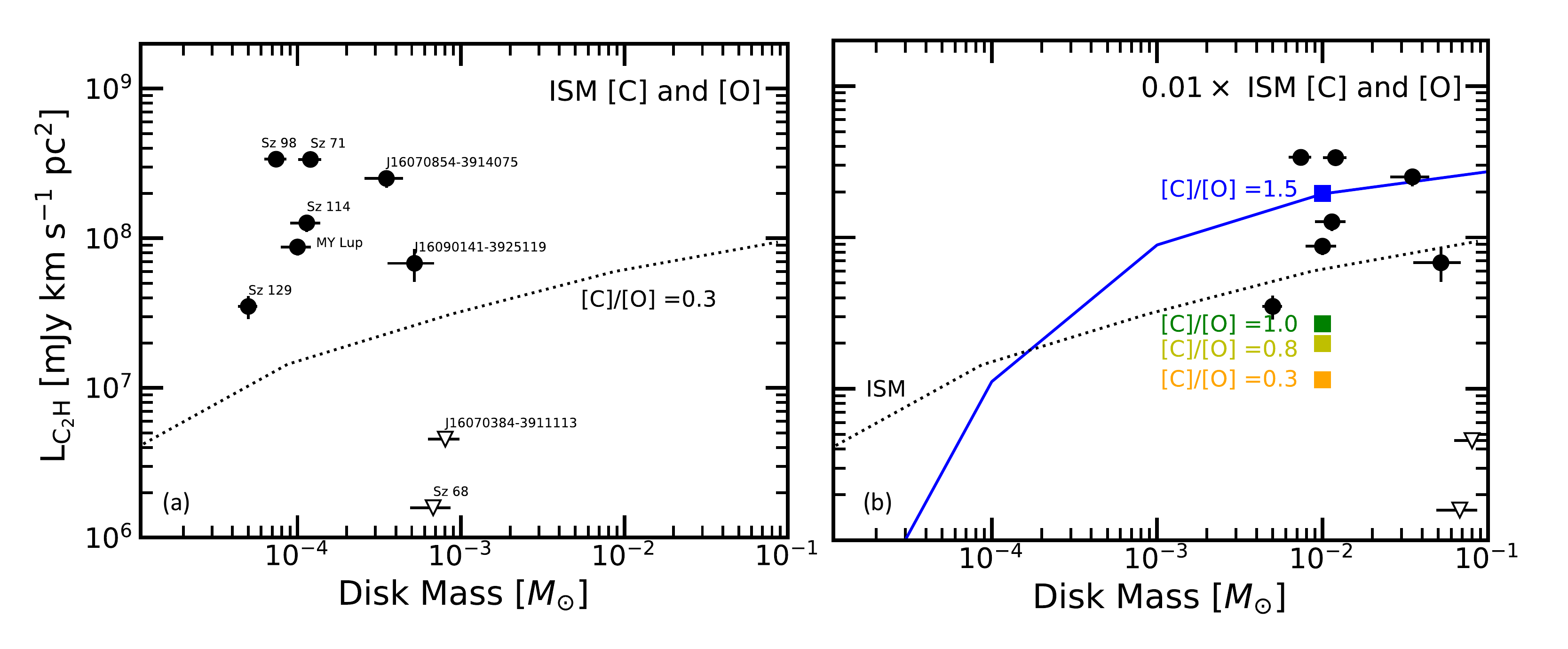}}
      \caption{Simulated C$_2$H integrated line luminosities using the code DALI \citep{Bruderer12} as functions of the disk gas masses compared with the Lupus observations. \emph{Panel (a):} the dotted black line shows model results obtained assuming [C] and [O] ISM-like abundances and [C]/[O]=0.3. Lupus detections are shown by the black circles, non-detections are shown by the empty triangles as 3-$\sigma$ upper limits. \emph{Panel (b):} the solid blue line shows model results obtained assuming that [C] and [O] abundances are depleted by two orders of magnitudes and [C]/[O]=1.5. Orange, light green, green, and blue squares show the model predictions when elemental [C] and [O] are depleted by two orders of magnitude and volatile [C]/[O]=0.3, 0.8, 1.0, and 1.5, for $M_{\rm disk}=10^{-2}M_{\odot}$. The black circles and empty triangles show C$_2$H observations as a function of CO-based gas masses, multiplied by a factor of 100, to account for the carbon depletion. }
       \label{DALI}
\end{figure*}

The simulated C$_2$H integrated line luminosities are presented in Figure \ref{DALI} as functions of gas disk mass and compared with the Lupus observations. In panel (a) the model results obtained with ISM-like gas-phase elemental abundances are shown by the dotted black line, while the observed C$_2$H integrated line luminosities are shown by the black symbols vs the CO-based gas masses. Our model results underestimate the detected C$_2$H integrated line luminosities shown by the black circles by one to two orders of magnitude. 

A subsample of models has been run reducing the carbon abundance by two orders of magnitude and the oxygen abundance by slightly larger factors, in order to reach volatile [C]/[O]=0.3, 0.8, 1.0 and 1.5. The model results, presented in panel (b) of Fig. \ref{DALI}, are shown by the coloured squares (in orange, light green, dark green, and blue for [C]/[O]=0.3, 0.8, 1.0 and 1.5 respectively for $M_{\rm disk}=10^{-2} M_{\odot}$).  A set of models with [C]/[O]=1.5 have been run for the whole mass range ($M_{\rm disk}=10^{-5}-10^{-2} M_{\odot}$) and the simulated C$_2$H integrated line luminosities are shown by the blue solid line. If the global volatile carbon and oxygen abundances are decreased by two orders of magnitude, but the [C]/[O] ratio is kept fixed to 0.3, the C$_2$H integrated line luminosities are fainter by a factor of five (see orange square in comparison with the dotted black line). If however the volatile [C]/[O] ratio increases and overcomes the  [C]/[O]=1 threshold, C$_2$H line luminosities are boosted by more than one order of magnitude (see comparison between orange and blue squares). Our Lupus C$_2$H detections may be compatible with a volatile carbon depletion of two orders of magnitude, and a [C]/[O]>1, as shown in panel (b) where the observed data points are shifted toward larger disk masses by a factor of 100. If carbon is depleted by two orders of magnitude, CO isotopologues lines would also be fainter by roughly the same factor and the actual disk masses would be a factor of 100 larger \citep{Miotello16}. This first order comparison with model predictions suggests that volatiles may be strongly depleted and [C]/[O]>1 in the Lupus disks. The two sources that were not detected in C$_2$H, Sz 68 and J16070384-3911113, may be disks with genuine gas depletion. In fact, according to Fig. \ref{DALI}, they would need to have very low masses to be reproduced by any of the models.

While our C$_2$H model results show that C$_2$H line emission increases by more than one order of magnitude if the volatile [C]/[O] ratio increases from 0.3 to 1.5 (see Fig. \ref{DALI}), the CN line fluxes, on the other hand, are found to vary only by a factor of 2 \citep[see][Fig. 14]{Cazzoletti18}. The stronger dependence of  C$_2$H lines on volatile [C]/[O] compared with CN lines may explain the steep correlation shown in Fig. \ref{CN} and tells us something about the similar but different origin of these two molecules.

\cite{Bergin16} investigated the effects of the volatile [C]/[O] on the intensity of the C$_2$H emission. The authors ran different sets of models with ISM-like volatile [C]/[O]=0.4, and with enhanced [C]/[O] ratios equal to 1 and 2. With an ISM-like abundance of carbon and oxygen ([C]/[O]=0.4) their simulated emission is well below observed values in DM Tau and TW Hya. With [C]/[O]=1 the simulation comes closer to the data, but [C]/[O]>1 is needed to approach the observed values, similarly to the Lupus disks. The same conclusion was found by \cite{Kama16} for TW Hya. This shows that bright C$_2$H observations do not simply imply a large volatile [C]/[O] ratio, but specifically that such ratio is larger than unity.  
Similarly, two orders of magnitude increase in the C$_2$H simulated column densities is found by \cite{Cleeves18} (see their Fig. 4) going from [C]/[O]=0.95 to 1.86. A similar behaviour is found also with our DALI models, where C$_2$H column densities increase by two orders of magnitude if we increase [C]/[O] from 0.8 to 1.5 (see Fig. \ref{N}).

\begin{figure}[h]
   \resizebox{\hsize}{!}
             {\includegraphics[width=2\textwidth]{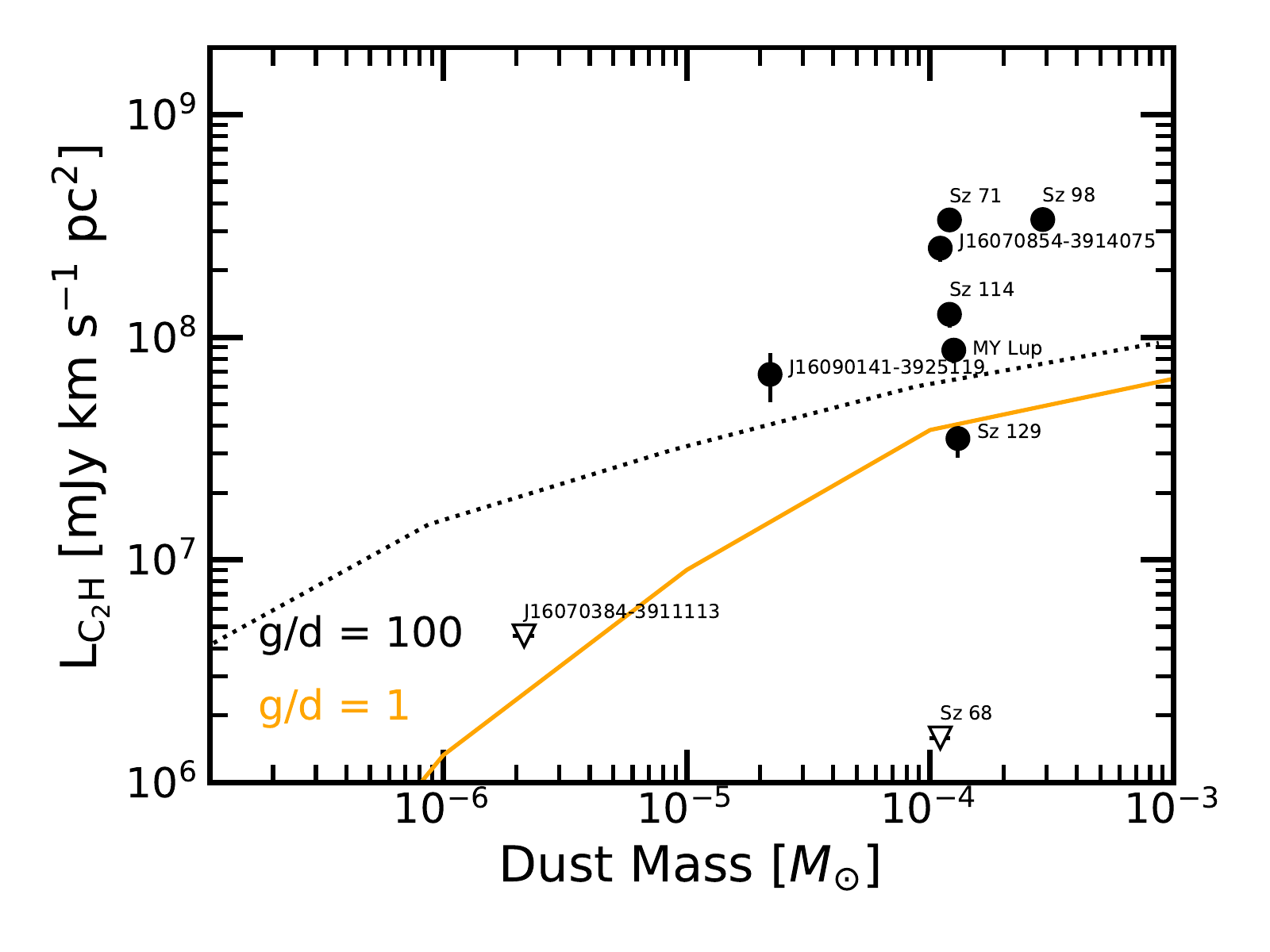}}
      \caption{Simulated C$_2$H integrated line luminosities using the code DALI \citep{Bruderer12} as functions of the disk dust masses compared with the Lupus observations. The dotted black line shows model results obtained assuming [C] and [O] ISM-like abundances and a gas/dust ratio of 100 (same as in panel a of Fig. \ref{DALI}). The solid orange line shows model results obtained decreasing the gas masses by two orders of magnitude, while keeping the dust mass fixed, in order to obtain a gas/dust ratio of 1. Lupus detections are shown by the black circles, non-detections are shown by the empty triangles as 3-$\sigma$ upper limits. }
       \label{DALI_2}
\end{figure}

In order to test the dependence of the C$_2$H emission with an actual gas depletion, we ran the same set of standard models shown by the dotted black line in Fig. \ref{DALI}, but decreasing the gas masses by two orders of magnitude while keeping the same dust masses. This provides a set of models with gas/dust ratios equal to unity, whose results are presented in Fig. \ref{DALI_2} by the orange solid line. Our Lupus C$_2$H observations are also reported, as a function of their dust masses, derived from their continuum emission \citep[see][]{Miotello17}. These models fail to reproduce the brighter observed C$_2$H line luminosity even more than the models with a gas/dust ratio of 100. The fainter C$_2$H detection, i.e. Sz 129, could be consistent with models with gas/dust=1 if we compare with the disk dust mass (Fig. \ref{DALI_2}), but this does not hold if we plot the model results against the total gas mass based on CO for ISM [C] and [O] (as in Fig. \ref{DALI}, left). Interestingly enough, J16070854-3914075 may be compatible with a genuine gas depletion. Sz 68 is a more complicated case, as part of a known binary system \citep{Andrews18}.

\section{Summary and conclusion}

This work presents new ALMA observations of C$_2$H together with higher signal/noise continuum data at 246 GHz in a subsample of nine protoplanetary disks in the Lupus star-forming region. Seven out of nine disks are detected in C$_2$H emission. After performing self-calibration, the C$_2$H integrated luminosities have been compared with CO-based gas/dust ratios \citep{Miotello17}. The seven detections have much brighter fluxes than the typical sensitivity of our data, hinting to a step-like distribution of the disks in our sample that results from the [C]/[O] ratio being larger or smaller than unity.  This interpretation is supported by physical-chemical models, which are able to reproduce the observations only if volatile carbon and oxygen are depleted and the [C]/[O] ratio is larger than unity. Models with reduced gas/dust ratios fail instead to reproduce our C$_2$H observations. The small number statistics however limits the study, but the conclusion is strengthened when literature data on other sources are included. Other effects such as UV luminosity and disk size may cause scatter from source to source.

A steeper than linear correlation is found between C$_2$H and CN emission lines in our Lupus disks. This can be explained by the fact that C$_2$H line fluxes depend more strongly on variations of the volatile [C]/[O] ratio than CN line fluxes. This implies that these two molecules have a similar but different origin in disks.

Ring-like structures are detected both in C$_2$H emission and in continuum \citep{Huang18} but they do not seem to be connected. Sz 71 shows ring shaped emission both in C$_2$H and in CN and the location of the peak intensity coincides for the two tracers.

The results presented in this paper suggest that volatile carbon and oxygen depletion is at play in the Lupus disks, and it is consistent with the $^{13}$CO and C$_2$H observations. This however does not necessarily rule out fast gas dispersal (see e.g., J16070854-3914075), as an explanation for the low CO fluxes and CO-based gas masses, since actual gas disk masses remain difficult to quantify \citep{Ansdell16,Pascucci16,Miotello17,Long17}. This has important implications for disk evolution and planet-formation theories, as disk gas masses are possibly larger than what CO observations would suggest if CO is assumed to be the main carbon carrier in the gas phase.

\section*{Acknowledgements}

The authors wish to thank Richard Booth, Antonella Natta, Karin {\"O}berg, and Jennifer Bergner for insightful comments and interesting discussions. A.M. and S.F. acknowledge an ESO Fellowship. This project has received funding from the European Union's Horizon 2020 research and innovation programme under the Marie Sklodowska-Curie grant agreement No 823823.  This work was funded by the Deutsche Forschungsgemeinschaft (DFG, German Research Foundation) - Ref no. FOR 2634/1 ER685/11-1. This paper makes use of the following ALMA data: ADS/JAO.ALMA\#2017.1.00495.S. ALMA is a partnership of ESO (representing its member states), NSF (USA) and NINS (Japan), together with NRC (Canada), MOST and ASIAA (Taiwan), and KASI (Republic of Korea), in cooperation with the Republic of Chile. The Joint ALMA Observatory is operated by ESO, AUI/NRAO and NAOJ.



\begin{appendix}
\section{Dependence of C$_2$H and CN emission on $L_{\star}$ and $L_{\rm acc}$.}
\label{appendix}

\begin{figure}[h]
   \resizebox{\hsize}{!}
             {\includegraphics[width=2\textwidth]{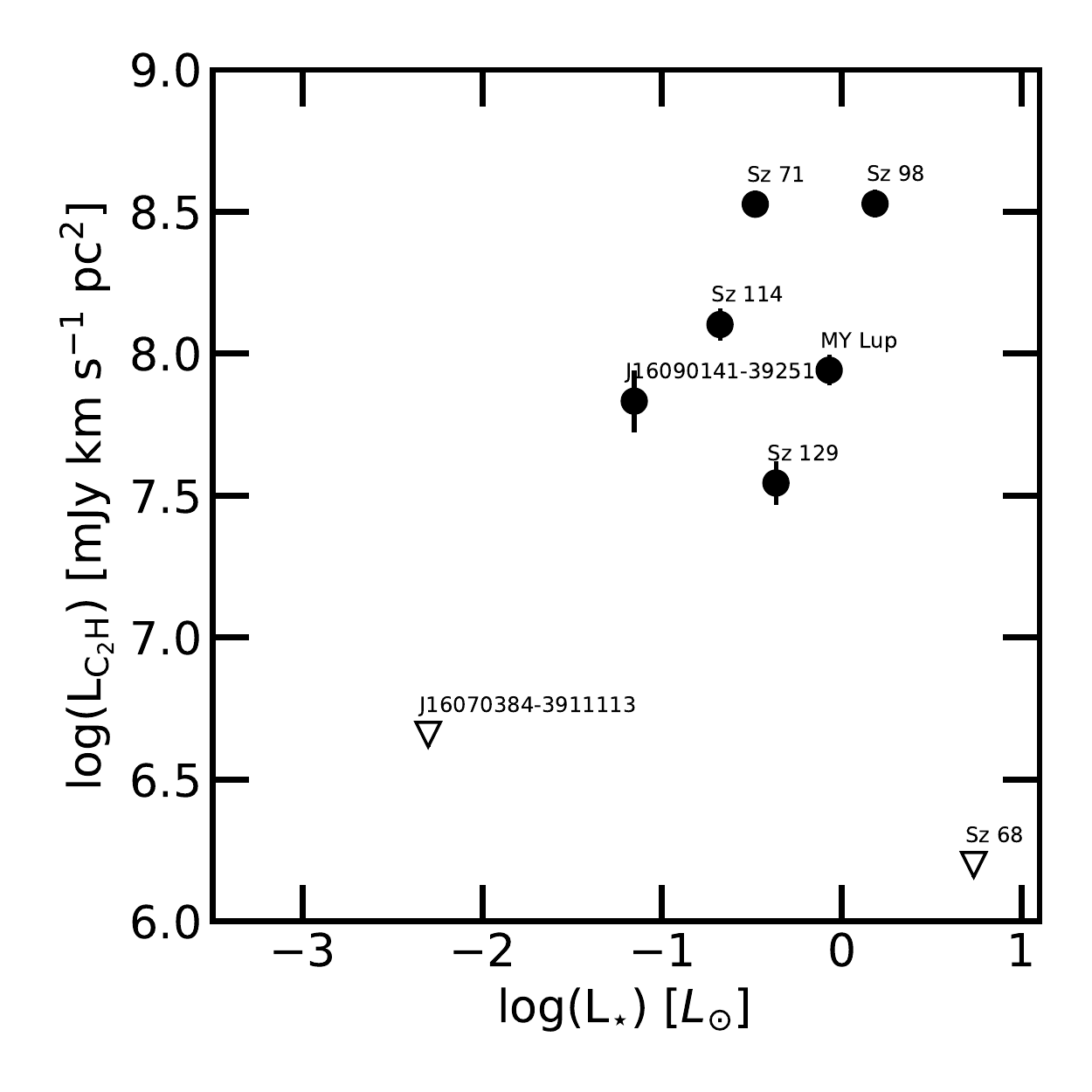}}
      \caption{Logarithm of C$_2$H ($N=3-2, J=7/2-5/2, F=4-3$ and $F=3-2$) versus stellar luminosity \citep{Manara18} for the Lupus disks. Empty triangles show the C$_2$H non-detections as 3-$\sigma$ upper limits.}
       \label{C2H_Lstar}
\end{figure}

\begin{figure}[h]
   \resizebox{\hsize}{!}
             {\includegraphics[width=2\textwidth]{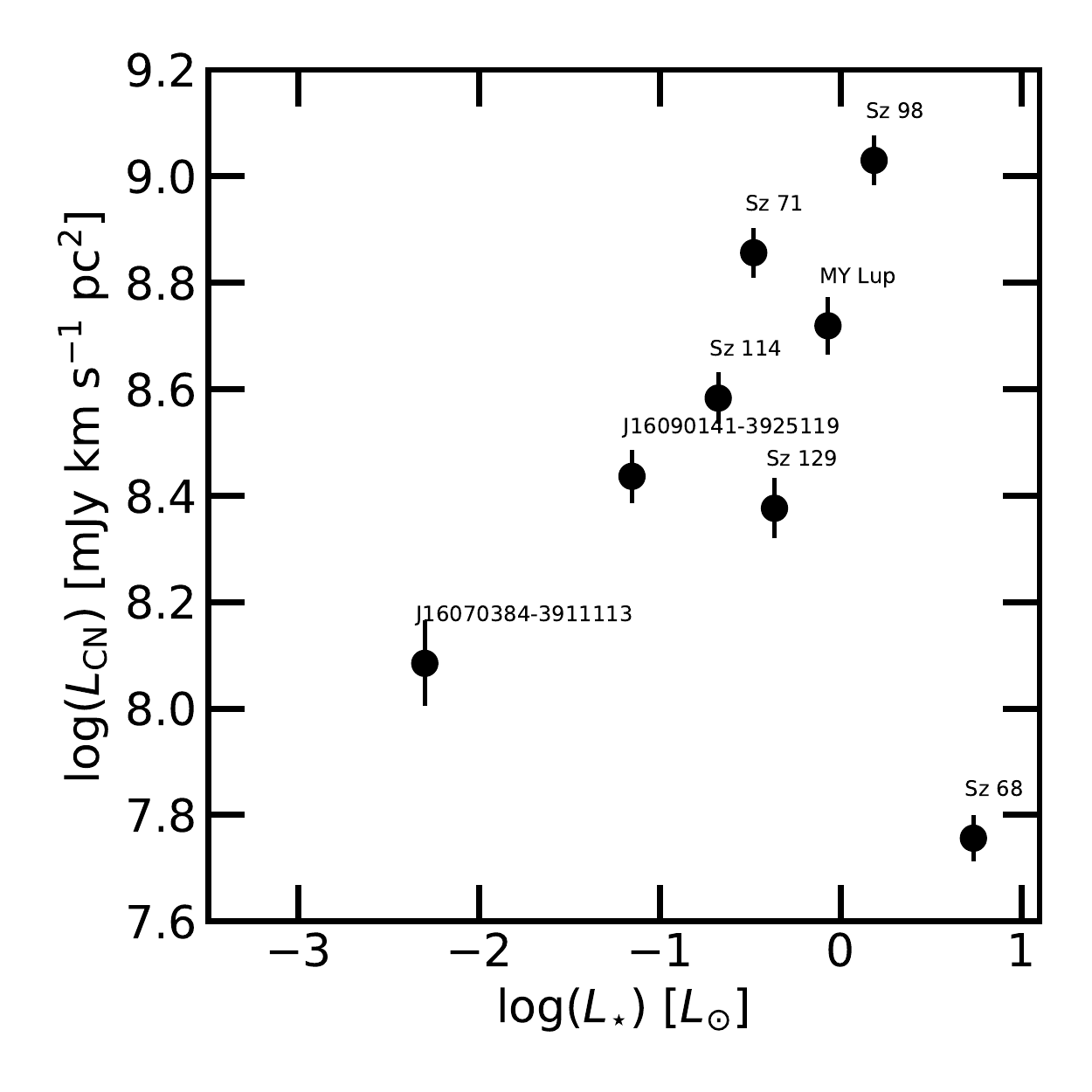}}
      \caption{Logarithm of CN (N=3-2) versus stellar luminosity \citep{vanTerwisga18,Manara18} for the Lupus disks.}
       \label{CN_Lstar}
\end{figure}

\begin{figure}[h]
   \resizebox{\hsize}{!}
             {\includegraphics[width=2\textwidth]{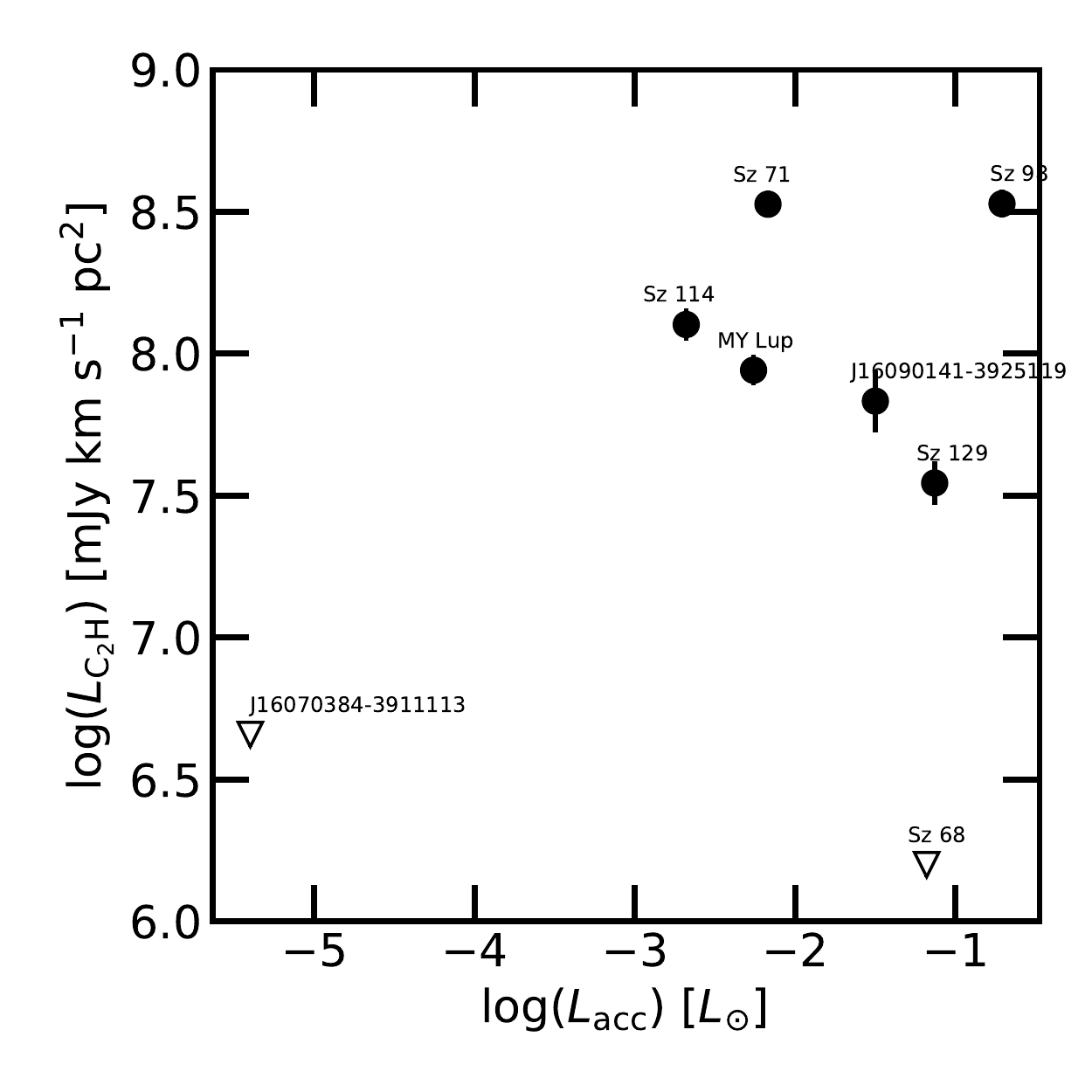}}
      \caption{Logarithm of C$_2$H ($N=3-2, J=7/2-5/2, F=4-3$ and $F=3-2$) versus accretion luminosity \citep{Manara18} for the Lupus disks. Empty triangles show the C$_2$H non-detections as 3-$\sigma$ upper limits.}
       \label{C2H_Lacc}
\end{figure}

\begin{figure}[h]
   \resizebox{\hsize}{!}
             {\includegraphics[width=2\textwidth]{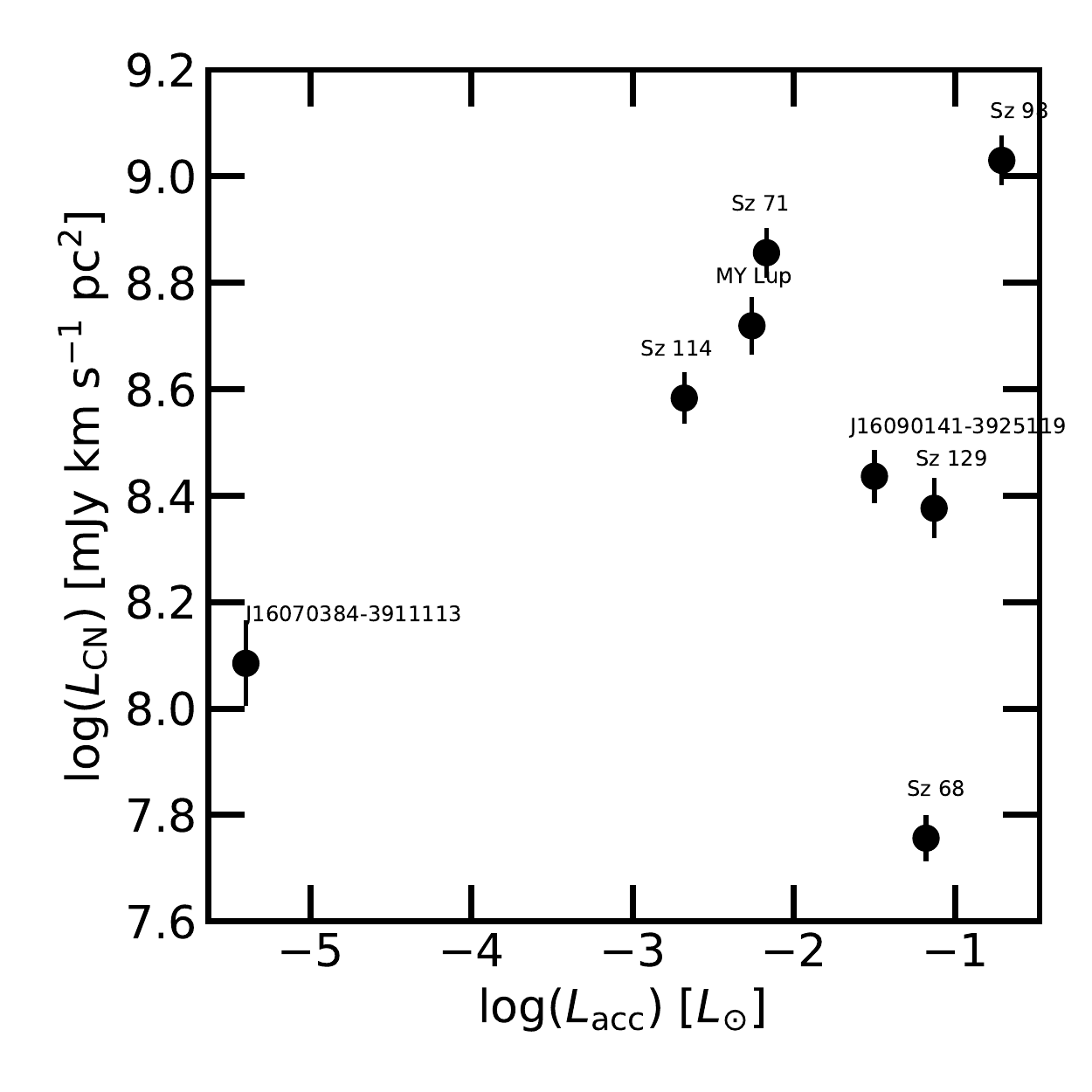}}
      \caption{Logarithm of CN (N=3-2) versus accretion luminosity \citep{vanTerwisga18,Manara18} for the Lupus disks.}
       \label{CN_Lacc}
\end{figure}

The nine Lupus disks targeted in this paper have also been detected by \cite{vanTerwisga18} in CN ($N=4-3$) emission. In this section we plot C$_2$H and CN emission line luminosity as a function of stellar and accretion luminosity \citep[][see Tab. \ref{Tab1}]{Alcala17} in Fig. \ref{C2H_Lstar}, \ref{CN_Lstar}, \ref{C2H_Lacc}, \ref{CN_Lacc}.


\newpage
\section{C$_2$H column densities.}
\label{appendixb}

\begin{figure}[h]
   \resizebox{\hsize}{!}
             {\includegraphics[width=2\textwidth]{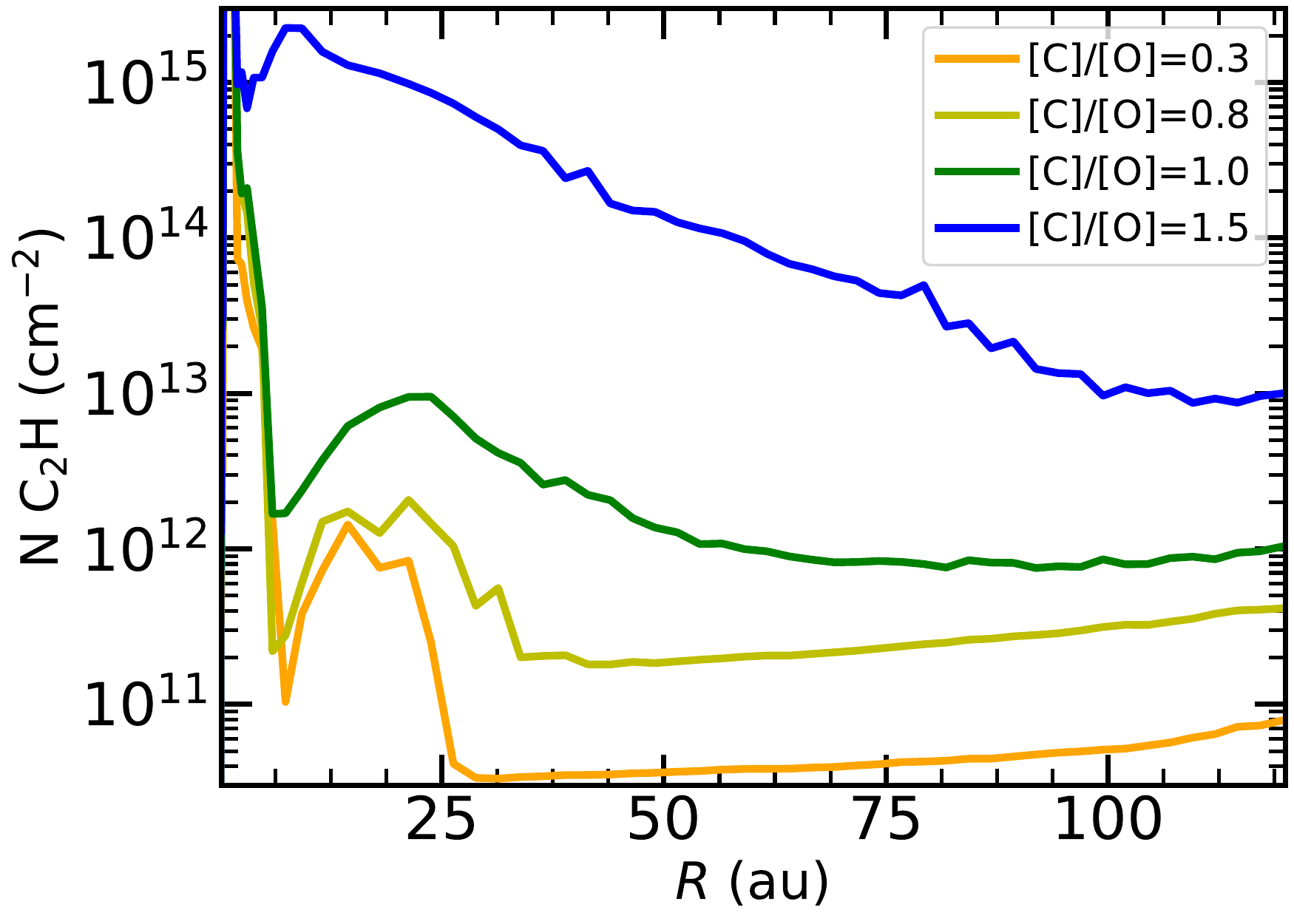}}
      \caption{C$_2$H column densities as a function of the disk radius obtained with DALI for a 10$^{-2}M_{\odot}$ disk. Volatile C abundance is reduced by two orders of magnitude and O abundance is decreased by the same value or more in order to obtain [C]/[O] ratios of 0.3 (orange line),  0.8 (light green line), 1.0 (green line) and 1.5 (blue line).}
       \label{N}
\end{figure}

The simulated C$_2$H column densities obtained with a subsample of our DALI models are shown in Fig. \ref{N}. 
The total disk mass is fixed to 10$^{-2}M_{\odot}$, volatile C abundance is reduced by two orders of magnitude (with respect to ISM abundance) and O abundance is decreased by the same value or more in order to obtain [C]/[O] ratios of 0.3 (orange line),  0.8 (light green line), 1.0 (green line) and 1.5 (blue line). Similarly to what is found by \cite{Cleeves18} (see their Fig. 4), the C$_2$H column density increases by more than one order of magnitude if we increase [C]/[O] from 0.8 to 1.5 (see FIg. \ref{N}).

\end{appendix}
\end{document}